\documentclass{elsarticle}
 \pdfoutput=1
\usepackage[utf8]{inputenc}
\usepackage{amsmath}
\usepackage{amsfonts}
\usepackage{amssymb}
\usepackage{graphicx}
\usepackage{color}
\usepackage{url}
\begin{document}
\title{Interband excitations in the 1D limit of two-band fractional Chern insulators}
\author[pwr]{B\l a\.{z}ej Jaworowski\corref{cor1}}
\ead{blazej.jaworowski@pwr.edu.pl}
\author[pwr]{Piotr Kaczmarkiewicz}
\ead{piotr.kaczmarkiewicz@pwr.edu.pl}
\author[pwr]{Pawe\l ~Potasz}
\ead{pawel.potasz@pwr.edu.pl}
\author[pwr]{Arkadiusz W\'{o}js}
\ead{arkadiusz.wojs@pwr.edu.pl}
\address[pwr]{Department of Theoretical Physics, Wroc\l aw University of Science and Technology, Wybrze\.{z}e Wyspia\'{n}skiego 27 50-370 Wroc\l aw, Poland}
\cortext[cor1]{Corresponding author}
\begin{abstract}We investigate the stability of the one-dimensional limit of $\nu=1/3$ Laughlin-like fractional Chern insulator with respect to the interband interaction. We propose a construction for the excitations in the infinite-interaction case and show that the energy gap remains finite in the thermodynamic limit. Next, by means of exact diagonalization and Density Matrix Renormalization Group approaches, we consider deviations from ideal dimerization and show that they reduce the stability of the FCI-like states. Finally, to show that our approach is not restricted to one model, we identify the dimer structure behind the thin-torus limit of other system -- the checkerboard lattice.
\end{abstract}

\maketitle
\section{Introduction}
In recent years, the possibility of realization of quantum Hall effect in lattice systems has been intensely studied. The idea can be traced back to the work of Thouless, Kohomoto, Nightingale and den Nijs \cite{TKNN}, who have proven that the quantized Hall conductance in integer quantum Hall effect is proportional to a topological invariant, an integer named the Chern number, which can be assigned to every isolated energy band in a periodic system. For Landau levels its absolute value is 1. A similar result can be obtained in a crystal: an energy band can be characterized by a nonzero Chern number and therefore behave like a Landau level, i.e.\ exhibit nonzero quantized Hall conductivity. In the tight binding formalism, such a crystal would require complex hopping integrals. Haldane has shown that they can be induced by a pattern of magnetic field which is zero on average \cite{Haldane}. Such systems were named Chern insulators. Later, they were realized experimentally, entirely without a magnetic field, with the complex hoppings induced by the spin-orbit interaction \cite{ChernExperiment} or by a periodic driving of optical lattices \cite{ChernExperiment2}.  

These ideas were later extended to the fractional quantum Hall effect (FQHE) \cite{tsui, laughlin}. Analogs of fractional quantum Hall states, named fractional Chern insulators (FCIs), were theoretically predicted to exist in lattice systems whose energy bands are not only topologically nontrivial (i.e. characterized by nonzero Chern number), but also nearly flat \cite{Tang, Neupert, SunNature, PRX, RevParameswaran, RevBergholtz, RevNeupert}. Hence, the interaction dominates over the single-particle energy (in analogy to kinetic energy quenching in Landau levels), which is crucial for emergence of strongly correlated phases such as FCIs. A number of ways to realize them experimentally was proposed, including optical lattices \cite{opticalflux1, dipolarspin1, dipolarspin2} and transition metal oxide heterostructures \cite{NatureTransMet} or layers \cite{layers1, layers2}, as well as polariton systems \cite{polaritons}, although none of them has been successfully implemented so far. Such a realization would be beneficial from the perspective of fundamental research. On the one hand they would be a potentially more convenient platform for study of FQHE systems, available with no magnetic field and in higher temperatures (see the discussion in Ref. \citenum{RevBergholtz}). On the other hand, the FCI physics is richer than the one of FQHE.  New phenomena can be studied, e.g. breaking the magnetic translation symmetry \cite{BernevigCounting, SingleParticleChamon, lee2013pseudopotential} or particle-hole symmetry \cite{SingleParticleGrushin, hierarchy}, and  FCI series on bands with Chern number higher than $C=1$ \cite{highchern1, highchern2, highchern3, highchern4, moller2015fractional}. Moreover, the non-Abelian FCI series \cite{PRX, MooreReadWang, highchern4} may find application in quantum information processing, as they may allow to construct a topological quantum computer. This includes also the new states on the bands with Chern number $C=2$, inequivalent to those known from FQHE \cite{highchern4}.

An intuitive understanding of fractional quantum Hall systems can be gained from the so-called thin-torus limit \cite{TaoThouless, bergholtz2006pfaffian, BergholtzTT}, also known as Tao-Thouless (TT) limit. In this approach, one considers a Landau level on a torus with one of its circumferences tending to zero (alternatively, one may view it as neglecting all the interaction matrix elements other than density-density ones). The FQHE states are then adiabatically deformed into charge density wave (CDW) states \cite{BergholtzTT, rezayi1994laughlin, nakamura2011beyond}. Application of similar approach to FCI was considered in several works \cite{bernevig2012thin, guo2012fractional, budich2013fractional, grusdt2014realization, wang2013tunable,xu2013fractional}. In particular, it was found \cite{bernevig2012thin, guo2012fractional, budich2013fractional} that the 1D limit of the two-orbital Chern insulator model can be mapped into the Su-Schriefer-Heeger (SSH) model of polyacetylene \cite{su1979solitons, heeger1988solitons}, and in the fully dimerized case (corresponding to exactly flat bands) the CDW ground states can be obtained analytically \cite{bernevig2012thin, budich2013fractional}. However, it was shown \cite{budich2013fractional} that they are no longer FCIs, because they belong to a different class of topological phases. Instead of being topological orders like FCIs, they are symmetry-protected topological phases, since the former are not possible in strictly 1D systems with conserved particle number \cite{chen2010local, chen2011complete}. However, they retain some properties of the ``parent'' FCI states, such as the degeneracy and the momentum counting, as well as the spectral flow \cite{guo2012fractional} (while lacking others, such as the counting of states in entanglement spectrum \cite{bernevig2012thin, budich2013fractional}).

One of the fundamental issues in FCI research is finding the conditions of its stability. The topological flat bands are not exact analogs of Landau levels, and the numerical calculations show that at some values of model parameters the FCIs do not exist even though the single-particle bands are flat and nontrivial \cite{zoology}. Some conditions of their stability were specified, based on the flatness of Berry curvature \cite{BernevigCounting, BerryParameswaran, SingleParticleMurthy, zoology} and quantum distance \cite{FubiniRoy, FubiniStudy, FubiniNeupert}. Also, the pseudopotential formalism for FCIs was formulated \cite{wu2013bloch, lee2013pseudopotential, lee2014lattice}, allowing for the systematic study of the effect of interaction form. These factors are connected with properties of a single band. A separate question is the influence of other bands. Initial estimation of the stability conditions for FCIs was the following: the interaction energy scale should be much larger than the single-band dispersion, but smaller than the inter-band energy gap, since the band mixing may be detrimental to these states. Later numerical calculations have shown that this is not necessarily true, as the 1/3 Laughlin FCIs in some two-band systems, are stable even for infinite interaction \cite{FarExceed}. However, since these results were purely numerical, we do not know what details of the band wavefunctions are responsible for this stability.

In this work we try to shed some light on these results by considering the one-dimensional limit of FCI and study its stability with respect to the interband excitations. Our starting point is the fully dimerized SSH Hamiltonian, for which we obtain the excitation energies analytically and show that the energy gap remains finite even in thermodynamic limit. Next, we analyze the effect of inter-dimer hopping and staggered potential on the energy spectrum. We show that they decrease the energy gap and if their value is sufficiently high, they may eventually lead to destabilization of the FCI-like states. Finally, we interpret the 1D limit of checkerboard model in terms of dimerized wavefunctions, suggesting that our approach is valid also outside the two-orbital model.

\begin{figure}
\begin{center}
\includegraphics[width=\textwidth]{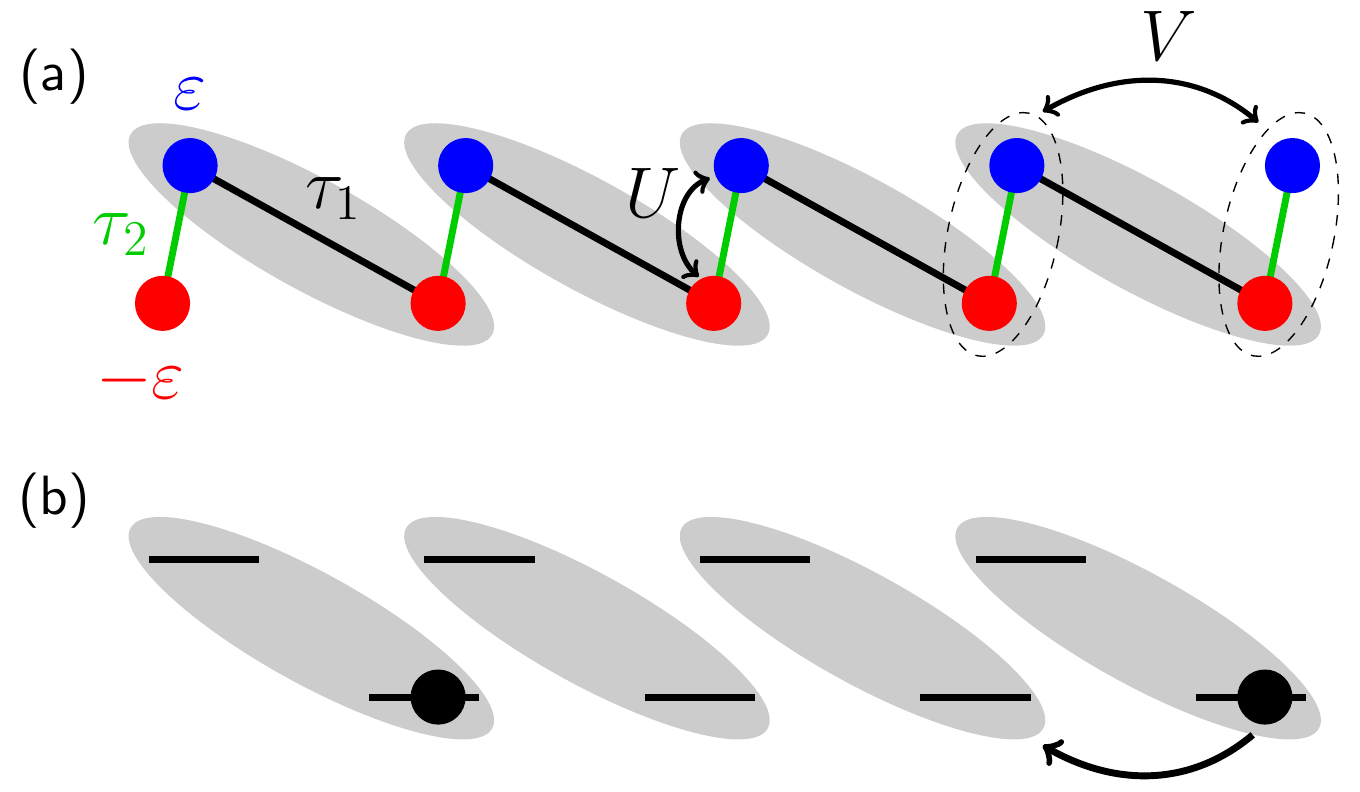}
\end{center}
\caption{(a) The extended SSH model. Each unit cell of the 1D chain contains two orbitals $\mathrm{A}$, $\mathrm{B}$ in the unit cells (red and blue circles, respectively) with onsite energies $\pm\varepsilon$. The system can be understood as a chain of sites (dashed ellipses) or dimers (grey ellipses), with inter-site (intra-dimer) hopping $\tau_1$ and intra-site (inter-dimer) hopping $\tau_2$. $U$ is the many-body intra-site interaction between $\mathrm{A}$ and $\mathrm{B}$ orbitals and $V$ is the inter-site interaction between nearest neighbors. (b) The many-body ground state for filling factor $1/3$ of the fully dimerized SSH model ($\tau_2=0$). Single particle eigenstates $\gamma$ ($\delta$) are denoted by horizontal lower (upper) lines. The many-body ground state is formed by populating every third $\gamma$ state. An arrow represents a possible excitation created by moving one particle to a neighboring dimer.}
\label{fig:ssh}
\end{figure}

\section{The model}
The system under consideration is shown in Fig. \ref{fig:ssh}(a). It consists of a one dimensional chain of sites (dashed ellipses) with two orbitals $\mathrm{A}$ and $\mathrm{B}$ (red and blue circles, respectively). The most general Hamiltonian involving hoppings only within each site and between nearest neighbors is given by
\begin{equation}
\label{eq:genham}
H=\sum_{i}\Psi_i^{\dagger}E\Psi_i+\sum_{i}\left(\Psi_i^{\dagger}T\Psi_{i+1} + h.c.\right),
\end{equation}
where $E$ and $T$ are $2\times 2$ matrices and $\Psi^{\dagger}_i=[a^{\dagger}_i, b^{\dagger}_i]$, with $a^{\dagger}_i, b^{\dagger}_i$ being the creation operators corresponding to orbitals A and B. The extended SSH model \cite{su1979solitons, heeger1988solitons, bernevig2012thin, guo2012fractional, budich2013fractional} is given by the hopping matrices
\begin{equation}
\label{eq:matET}
E_{\mathrm{SSH}}=\left[\begin{matrix} -\varepsilon & \tau_2\\
\tau_2 & \varepsilon
\end{matrix}\right], ~~~~
T_{\mathrm{SSH}}=\left[\begin{matrix} 0 & 0\\
\tau_1 & 0
\end{matrix}\right],
\end{equation}
where $\tau_1, \tau_2$ are the hoppings between the A and B orbitals between neighboring sites and within a site, respectively, and $\varepsilon$ is the strength of staggered potential. In this work, we will consider mostly $|\tau_1|>|\tau_2|$. In such a case, it will be useful to introduce another way of understanding this model. We can regard it as a chain of dimers (gray ellipses), with intra-dimer hopping $\tau_1$ and inter-dimer hopping $\tau_2$. When $\tau_2=0$ we will call the model ``fully dimerized''.

 If $\varepsilon=0$ the model can host a topological phase protected by chiral symmetry \cite{ryu2010topological}. The topological phase transition occurs at $|\tau_1|=|\tau_2|$. If finite $\varepsilon$ is introduced, the chiral symmetry is broken, and the band gap does not close during the transition.
 
Different variants of the interacting SSH model were considered in Refs. \citenum{bernevig2012thin, guo2012fractional, budich2013fractional}. We choose the following interaction with onsite $U$ and intrasite nearest neighbor $V$ terms (see Fig. \ref{fig:ssh}(a))
\begin{equation}
\label{eq:interaction}
U\sum_{i}n_{\mathrm{A}i}n_{\mathrm{B}i}+V\sum_{i}n_{i}n_{i+1},
\end{equation}
where $n_{\mathrm{A}i}$ ($n_{\mathrm{B}i}$) is the particle density at orbital A (B) of site $i$, and $n_i=n_{\mathrm{A}i}+n_{\mathrm{B}i}$.

\subsection{SSH model -- the dimer basis}
In order to understand the ground state properties of this model, we consider the fully dimerized limit corresponding to $\tau_2=0$ and later we investigate coupling between dimers within the perturbation theory. The single-particle Hamiltonian can be diagonalized by switching to dimer basis
\begin{equation}
\label{eq:fullydimerized}
H_{\mathrm{dim}}=\tilde{\varepsilon} \sum_{i}(-\gamma^{\dagger}_i \gamma_i+\delta^{\dagger}_i \delta_i)
\end{equation}
where $\tilde{\varepsilon}=\sqrt{\varepsilon^2+\tau_1^2}$, and
$$
\gamma_i=\frac{1}{C}\left(\tau_1b^{\dagger}_i-\left(\sqrt{\tau_1^2+\varepsilon^2}+\varepsilon\right)a^{\dagger}_{i+1}\right),$$
$$
\delta_i=\frac{1}{C}\left(\left(\sqrt{\tau_1^2+\varepsilon^2}+\varepsilon\right)b^{\dagger}_i+\tau_1 a^{\dagger}_{i+1}\right),
$$
with normalization constant $C=\sqrt{2\varepsilon^2+2t^2+2\varepsilon\sqrt{t^2+\varepsilon^2}}$. When $\varepsilon=0$, these expressions reduce to the ones considered in Refs. \citenum{bernevig2012thin,guo2012fractional, budich2013fractional}. At the single-particle level, the system has two exactly flat 
bands with energies $\pm \tilde{\varepsilon}$. 

The many-body ground state can be easily constructed by projecting the interaction onto the lower band. In the basis of $\gamma$ wavefunctions (which are still labeled with a site index $i$ but now stretch over two sites $i$ and $i+1$), the $k$-th neighbor interaction becomes a $(k+1)$-th one. Therefore, the interaction in Eq. \ref{eq:interaction} becomes a second neighbor one after the projection. We consider the lowest band of the system with a filling factor $\nu=1/3$ and periodic boundary conditions. The interaction energy can be minimized by keeping at least two empty dimers between two filled ones (Fig. \ref{fig:ssh}(b)). This corresponds to a CDW state with $100100100\dots$ occupation pattern, with $0$ and $1$ denoting the empty and filled dimers, respectively. Two other ground states may be obtained by shifting this pattern by one ($010010010\dots$) or two dimers ($001001001\dots$), hence the ground state is three-fold degenerate, as for $\nu=1/3$ Laughlin FCI \cite{wen, HaldaneCounting, BernevigCounting}. We will say that this FCI is the parent FCI state of our thin-torus ground state. Although the FCI-like CDW state is obtained within a single-band projection, it is in fact the exact ground state of the system, as it minimizes both single-particle and interaction energies separately at the same time. 

For any Laughlin-like filling $\nu=1/q$ we can construct a similar $q$-fold degenerate ground state by choosing a $(q-2)$-th neighbor interaction. This construction is very similar to generalized Pauli principle in FCIs \cite{HaldaneCounting, BernevigCounting} although in the real space instead of the momentum space. Indeed, the counting of states of the thin torus limit of FCIs agrees with this principle for the ground state at $\nu=1/q$ and quasihole states \cite{bernevig2012thin}. On the other hand, FCIs on a torus satisfy this counting also for particle entanglement spectra \cite{ParticleEntanglement, PRX, BernevigCounting}, which is not the case in the thin-torus limit \cite{budich2013fractional, bernevig2012thin}. Moreover, as it was was pointed out in Ref. \citenum{budich2013fractional}, by an appropriate choice of interaction one can construct $q$-fold degenerate ground states also for even $q$'s. They do not have a parent FCI state, as fermionic FCIs cannot exist at these fillings. A similar situation arises in the thin-torus limit of FQHE, in which one can also form CDW ground states for any $\nu=1/q$, but only the ones with odd $q$ are adiabatically connected to FQHE state \cite{BergholtzTT}.

In addition to the ground states, we can consider the band-projected excitations, formed by moving one or more particles to the neighboring dimer (see Fig. \ref{fig:ssh}(b)). Hence, two or more filled dimers will be separated by one empty dimer only, which will yield a finite energy proportional to the $V$ term, due to the effective second-neighbor interaction. However, in contrast to the ground states, these excitations are not exact eigenstates of the system, which makes the band-projected picture is not sufficient. In this work we want to determine their energies taking into account both bands. We will check if the energy gap will still remain finite when interaction is increased to infinity, and investigate how the gap is affected by different perturbations.

\subsection{Relation to the FCI models}
The extended SSH model can be related to 1D limit of different 2D tight-binding models with nearly flat topological bands. This can be done by applying a basis transformation at each site defined by the matrix
$$
U(\phi)=\left[\begin{matrix}
\cos{\phi} & -\sin{\phi}\\
\sin{\phi} & \cos{\phi}
\end{matrix}\right].
$$
The Hamiltonian after rotation will have the same form as Eq. \ref{eq:genham}, but now with $\tilde{\Psi}_i=U(\phi)\Psi_i$, $\tilde{T}_{\mathrm{SSH}}=U(\phi)T_{\mathrm{SSH}}U(\phi)^{\dagger}$ and $\tilde{E}_{\mathrm{SSH}}=U(\phi)E_{\mathrm{SSH}}U(\phi)^{\dagger}$.

It was shown \cite{bernevig2012thin, guo2012fractional} that the SSH model is equivalent to the thin-torus limit of 2D two-orbital Chern insulator model, a spinless version of the model developed to describe the mercury telluride quantum wells \cite{bernevig2006twoorb} (see also Ref. \citenum{zoology} for additional description and many-body calculations). In the 1D limit, this model is defined by the the following hopping matrices
$$
T_{\mathrm{2orb}}=\left[\begin{matrix}
t_2 & -t\\
t & -t_2\end{matrix}\right]~~~
E_{\mathrm{2orb}}=\left[\begin{matrix}
-M' & 0\\
0 & M'\end{matrix}\right],
$$
where $M'$ is the staggered potential and $t$ ($t_2$) is the hopping integral between the same (different) orbitals of neighboring sites. $M'$ is related to the staggered potential $M$ of the 2D model as $M'=M-2t_2$. In Refs. \cite{bernevig2012thin, guo2012fractional} it was shown that this model is in fact the SSH model with $\varepsilon=0$, $\tau=2t$ and $\tau_2=-M'$, rotated by $\phi=\pi/4$. 

Note that Eq. \ref{eq:interaction} is independent on the basis rotation. Therefore, the mapping is valid also on the many-particle level, and the ground states described in the previous subsection are also the ground states of the 1D two-orbital model, although the dimer wavefunctions $\gamma, \delta$ will no longer be dimers after rotation.

Similarly, we relate the SSH model to the thin torus limit of the checkerboard model \cite{Sun, Neupert}. The Tao-Thouless limit of this model in the spinful case was studied in Ref. \citenum{liu2016realizing}, but we apply a different approach: we consider spinless particles, populating the system of width one instead of two unit cells. In this way, we can utilize the above approach, treating each checkerboard unit cell as one site of our chain. The mapping between such a model and the SSH model is only approximate, but it can become exact if we introduce an additional hopping $t_\varepsilon$, not present in the original model from Refs. \citenum{Sun} and \citenum{Neupert}.  Even without this modification, it can be shown that the checkerboard model after rotation has a strong intra-dimer hopping and weaker inter-dimer ones. This is true for the versions with and without third-neighbor hoppings included (Refs. \citenum{Sun} and \citenum{Neupert}, respectively). We describe the details of this correspondence in the \ref{sec:checkerboard}.

Thus, in this work we study the extended SSH model given by Eq. \ref{eq:genham} and Eq. \ref{eq:matET}, as a representative of a family of 1D limits of various 2D lattice models with topological flat bands. We focus on $\nu=1/3$, although the results will be easily generalizable to other fillings.

\section{Excitations in the fully dimerized system} \label{sec:fullydim}
We start from determining the excitation spectrum of the fully dimerized SSH model with no staggered potential ($\tau_2=0, \varepsilon=0$). In such a case the dimer occupation is conserved, which is a significant simplification in the construction of excitation wavefunctions. The one kind of excitations may be constructed by moving one particle forming the many-body ground state from $\gamma$ to $\delta$ eigenstate within a dimer. Its energy $E^{\mathrm{X}}_0$, measured relatively to the ground state, is therefore equal to the band gap $E^{\mathrm{X}}_0=2\tau_1$.
\begin{figure}
\includegraphics[width=\textwidth]{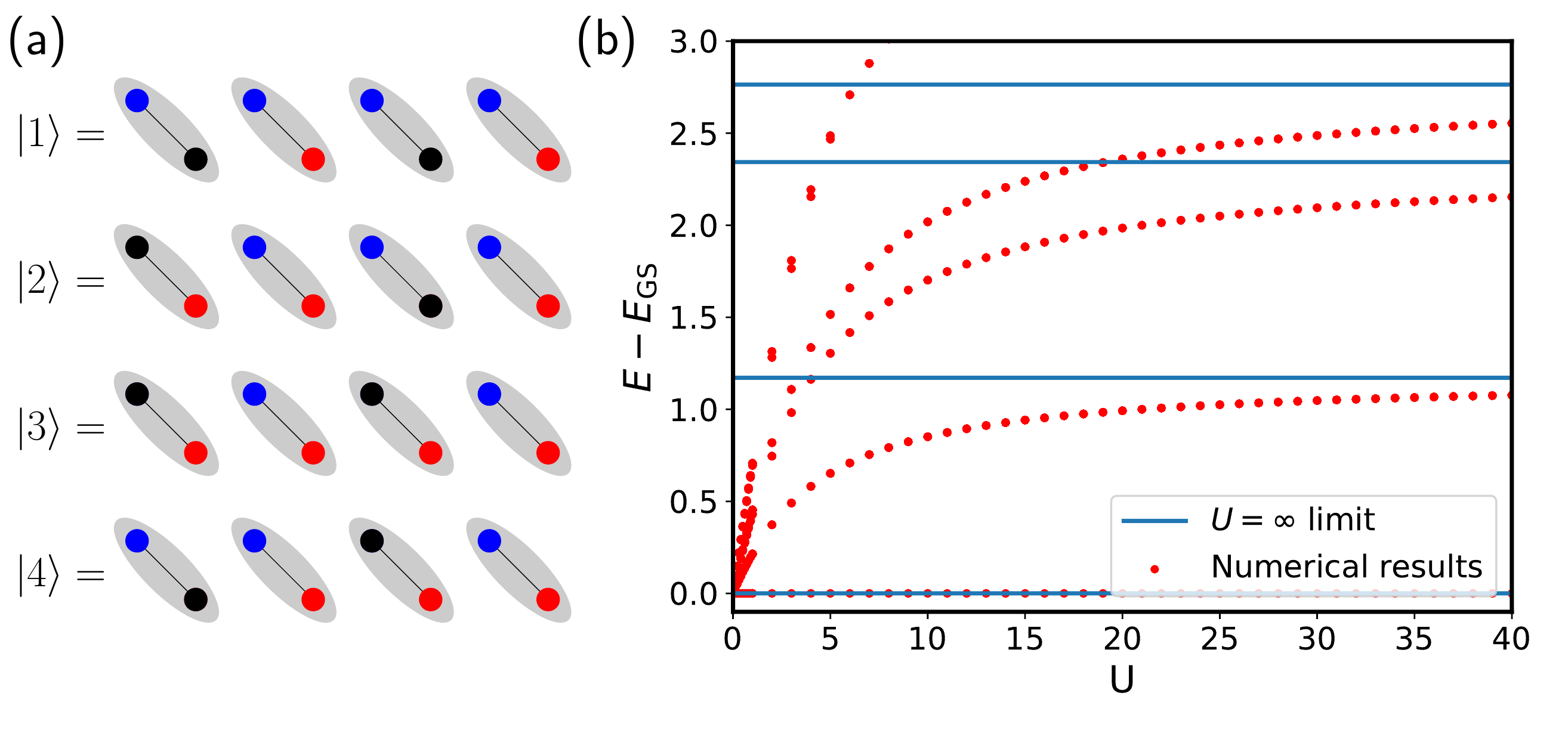}
\caption{The fully-dimerized limit of the SSH model. (a) Four configurations in one-particle excitation. Black circles denote filled A or B orbital. (b) Low-energy spectrum of a system with $N_{\mathrm{part}}=4$ particles on a chain of length $L=12$ sites as a function of interaction strength $U$, and $U=V$, with $\tau_1=2$. Blue lines denote the analytical result in an infinite-interaction limit given by Eq. \ref{eq:exc}.}
\label{fig:dimerized}
\end{figure}
Another type of excitation, which we will call a one-particle excitation, can be constructed by moving one of the particles in the ground state to the neighboring dimer, so that there is now only one empty dimer between the filled ones (Fig. \ref{fig:ssh}(b)). For example, starting from a ground state $1001001001\dots$ we can get an excitation with $1010001001\dots$ occupation pattern. The two particles involved in the excitation are isolated from the rest and interact only with each other, hence we can focus on the two dimers only and neglect the rest of the system in our analysis. We consider both bands, so there are four possible ways in which the particles can be distributed between the dimer states. It is convenient to define these four configurations in the basis of $\mathrm{A},\mathrm{B}$ orbitals belonging to the two dimers, rather than using the $\gamma, \delta$ wavefunctions. The configurations, listed in Fig. \ref{fig:dimerized}(a) are coupled to each other by the intra-dimer hopping integral $\tau_1$. The state $|4\rangle$ is the only one with a nonzero energy due to the $V$ term. We are interested in the limit $V\rightarrow\infty$ in which this energy is infinite, hence this state can be discarded from the basis. We are left with a $3\times 3$ effective Hamiltonian
\begin{equation}
H_{1}=\left[\begin{matrix}
0 & \tau_1 & 0 \\
\tau_1 & 0 & \tau_1 \\
0 & \tau_1 & 0 
\end{matrix}\right]
\label{eq:HX},
\end{equation}
whose lowest eigenvalue is $E_{1}=-\tau_1\sqrt{2}$. Using this result, we calculate the energy of these excitations with respect to the ground state. The three-fold degenerate ground state for the system with $N_{\mathrm{part}}$ particles has energy $E_{\mathrm{GS}}=-N_{\mathrm{part}}\tau_1$. The energy of the excitation, including all the remaining particles staying in the ground state, is $\tilde{E}_1=-\tau_1(N_{\mathrm{part}}-2)-\tau_1\sqrt{2}$. Hence, the energy with respect to the ground state is $E^{\mathrm{X}}_{1}=\tilde{E}_1-E_{\mathrm{GS}}=(2-\sqrt{2})\tau_1$. This energy level is highly degenerate, because we can choose different pairs of particles to create the excitation, and because we have some freedom of arranging the remaining particles without changing the energy.

In a similar way, we can consider $k$-particle excitation by moving $k$ particles: the first one by one dimer, the second by two dimers etc. As a result, $k+1$ particles are separated by one empty dimer only. For example, starting from $100100100100\dots$ ground state, we can obtain an excitation $101010000100\dots$ with $k=2$. The Hamiltonian of $k$-particle excitation will correspond to an open tight-binding chain of length $k+2$ with a tridiagonal matrix structure as in Eq. \ref{eq:HX}. The lowest energy of $k$-particle excitation energy with respect to the ground state can be written as
\begin{equation}
E^{\mathrm{X}}_k=(k+1)|\tau_1|-2|\tau_1|\cos\left(\frac{\pi}{k+3}\right)
\label{eq:exc}
\end{equation}
which is the lowest in the case of $k=1$. Hence $E_1$ determines the many-body energy gap in the fully dimerized case with large interaction for arbitrary number of particles, and also in the thermodynamic limit. We note that the excitation is localized, so the energy does not depend on the system size.

We compare the above analytical results with numerical ones. Fig. \ref{fig:dimerized}(b) shows the evolution of energy spectrum with increasing $U$ ($U=V$) for $N_{\mathrm{part}}=4$ particles on chain of $L=12$ sites obtained using the exact-diagonalization (ED) method. As $U$ increases, the energies converge to analytical results for infinite interactions, denoted by blue horizontal lines in the plot. The first energy level above the ground state correspond to the $k=1$ excitation. One can see that it does not intersect with other levels in the entire range of interaction strength, i.e. the $k=1$ excitation is the lowest one also for finite $U$.
\begin{figure}
\includegraphics[width=\textwidth]{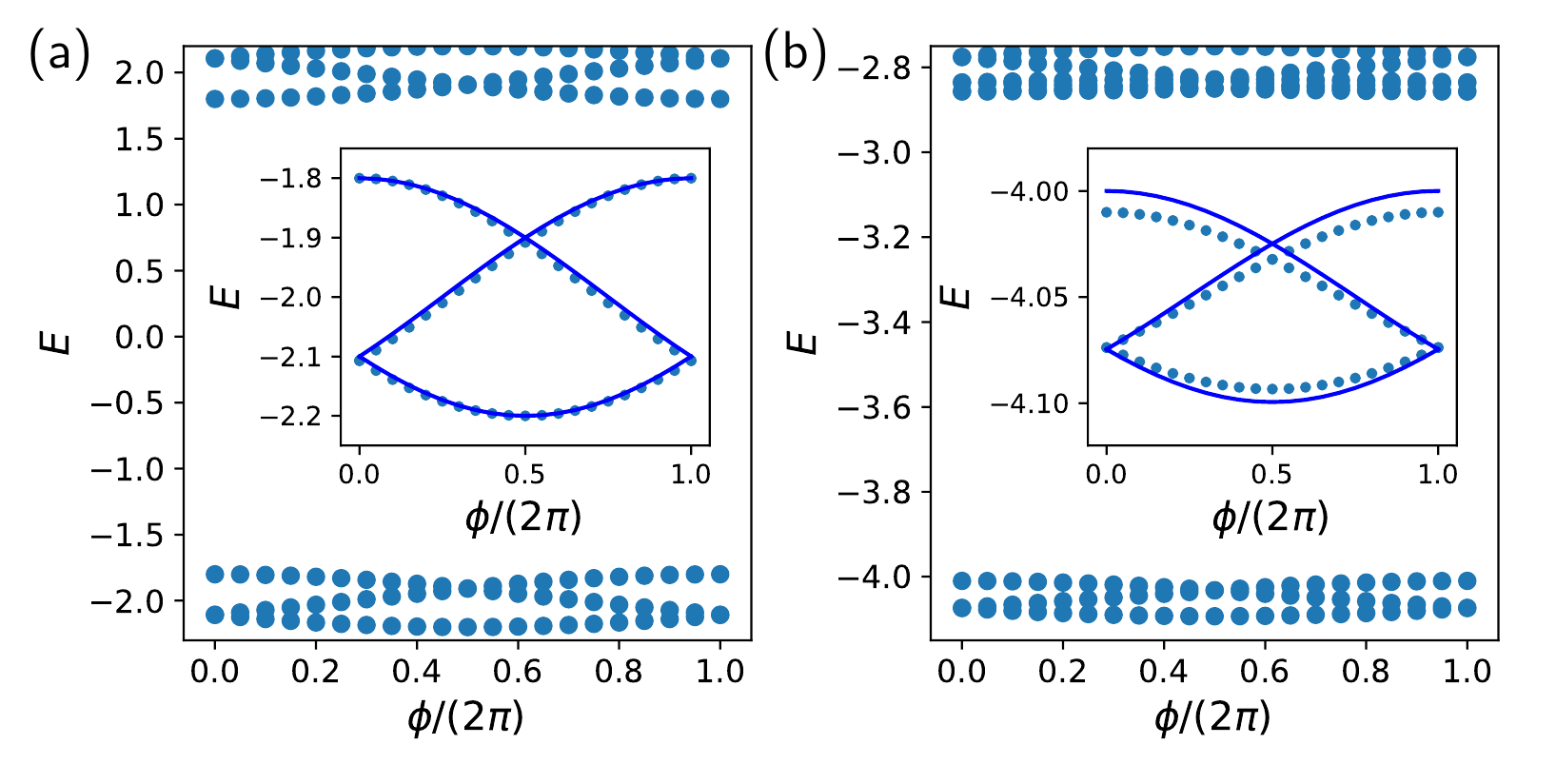}
\caption{Spectral flow in SSH chains with with $\tau_1=2, \tau_2=-0.2$, for (a) $L=3$, $N_{\mathrm{part}}=1$ (b) $L=6$, $N_{\mathrm{part}}=2$. The energies obtained using the exact-diagonalization method are labelled by dots. The insets show a closer view of three lowest states, with lines denoting the perturbation theory result.}
\label{fig:spectralflow}
\end{figure}

\section{Effects of the inter-dimer hopping}

\subsection{The ground state degeneracy removal}
We now turn on the inter-dimer hopping integrals between neighboring dimers $\tau_2$ in a perturbative way, assuming that $\tau_1\gg \tau_2$. The inter-dimer hoppings couple the three ground states to each other, lifting the perfect three-fold degeneracy. This allows one to observe the spectral flow when the twisted boundary conditions are applied \cite{guo2012fractional}. In FQHE (and hence FCI) the $q$ quasi-degenerate ground states flow into each other when the boundary phase is changed from $0$ to $2\pi$, and return to themselves when $2q\pi$ phase is reached\cite{sflow1}. Such a spectral flow is one of the characteristic features that allows to identify the FCI state (see e.g. Refs. \citenum{Neupert,zoology}), but it is not a definite proof that the system is an FCI, and should be complemented with other calculations (e.g. of the particle entanglement spectrum).

 We first consider systems consisting of $L=3$ sites with $N_{\mathrm{part}}=1$ particle, and $L=6$ sites with $N_{\mathrm{part}}=2$. The effect of $\tau_2$ can be derived from the perturbation theory of the first and second order, for one and two particle cases, respectively. We start from the three degenerate ground states of the fully dimierized system and introduce $\tau_2$ as a perturbation, via the Hamiltonian
$$H_{\mathrm{PT}}=\left[\begin{matrix}
B & A & A\exp(i\phi) \\
A & B & A \\
A\exp(-i\phi) & A & B
\end{matrix} \right],$$
where $A=\tau_2/2$, $B=0$ for $N_{\mathrm{part}}=1$, and $A=\frac{\tau_2^2}{8\tau_1(2-\sqrt{2})}$, $B=2A$ for $N_{\mathrm{part}}=2$, while $\phi$ is the boundary phase ($\phi=0$ for ordinary periodic boundary conditions and $\phi\neq 0$ for twisted ones). The eigenvalues of $H_{\mathrm{PT}}$ are $E_j=2 A \cos((2\pi j+\phi)/3)+B$ with $j=0,1,2$, i.e. the degeneracy is removed, and the three states flow into each other as $\phi$ is changed. In Fig. \ref{fig:spectralflow}, these results are compared to numerically obtained spectra, showing a good agreement with them. The occurrence of the spectral flow is consistent with the results in Ref. \citenum{guo2012fractional}, where the spectral flow was obtained numerically for a different kind of interaction of finite strength.

We note that these perturbation theory arguments can be applied to any filling $1/q$ (except $q=2$), so we are able to observe spectral flow for even fillings also. Therefore the notion of the spectral flow as a remnant of properties of the parent FCI state should be treated with caution, as for even $q$ there is no such state. 

The degeneracy splitting is smaller for $N_{\mathrm{part}}=2$ than for $N_{\mathrm{part}}=1$, see Fig. \ref{fig:spectralflow}(b) noting four times larger energy scale in the inset in comparison to Fig. \ref{fig:spectralflow}(a). We expect a further decrease with increasing $N_{\mathrm{part}}$, because we can obtain one ground state from the other only by moving all the particles, i.e. we have to use at least $N_{\mathrm{part}}$-th order of perturbation theory. Hence, one can expect that for small $\tau_2$ the degeneracy splitting of the ground states will vanish in a thermodynamic limit. 

\subsection{The infinite interaction limit}\label{ssec:infinite}
\begin{figure}
\includegraphics[width=\textwidth]{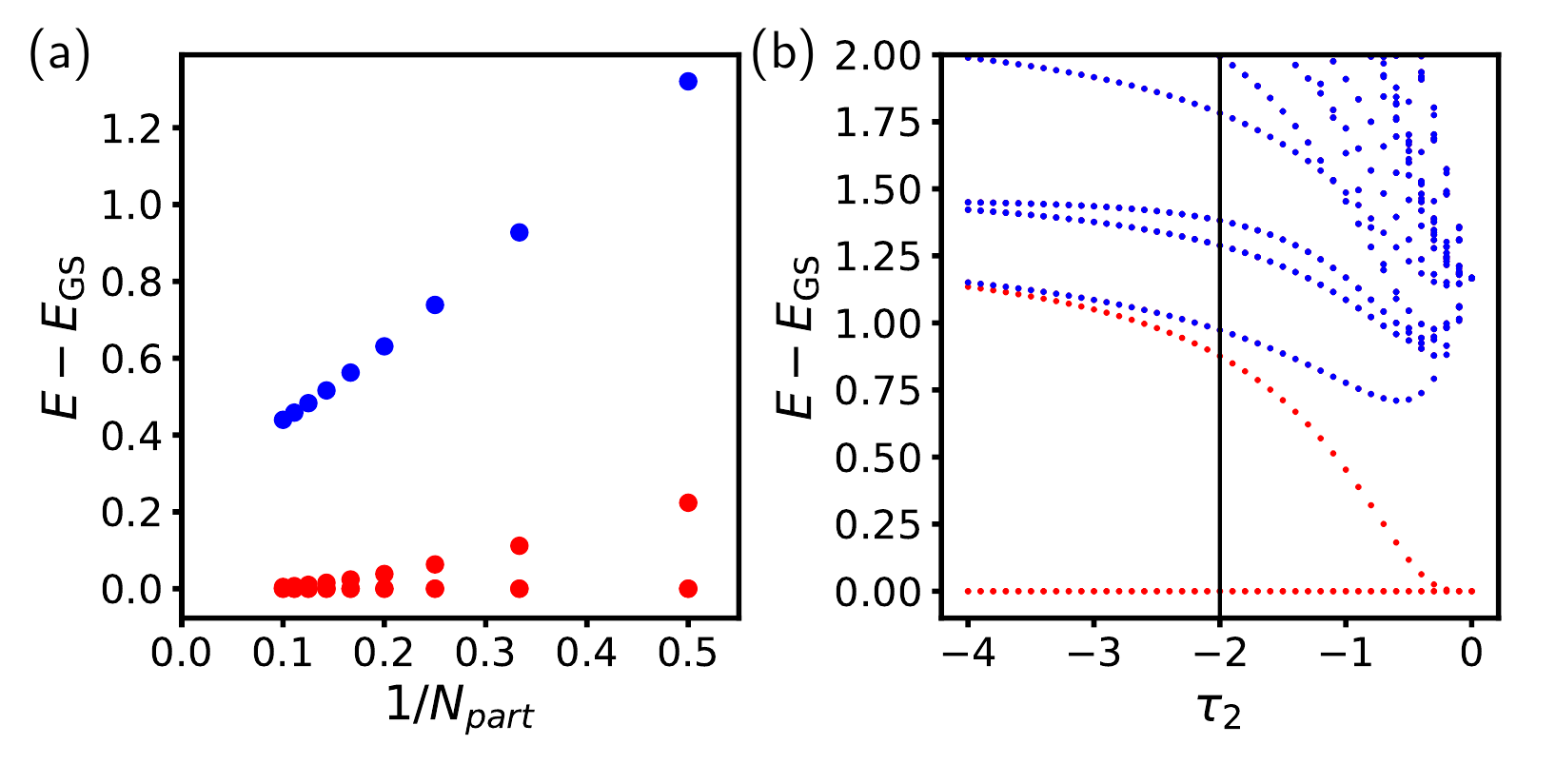}
\caption{Low-energy spectrum for the SSH model with $\nu=1/3$, $U=1000$, $\tau_1=2$ and finite $\tau_2$. (a) Scaling of the energy of four lowest states for $\tau_2=-0.4$ with inverse number of particles, obtained using DMRG method. (b) Low-energy spectrum of $L=12$ chain with $N_{\mathrm{part}}=4$ particles, obtained using the exact diagonalization method. Three lowest states are denoted by red points. The black line in (b) indicates the single-particle topological phase transition.}
\label{fig:dmrg}
\end{figure}

When finite $\tau_2$ is introduced, the degenerate $k=1$ excitations will couple to each other (as well as to other states). As a consequence, their degeneracy will be lifted, which will affect the energy gap. We study this effect using the DMRG method \cite{white1992density}. This approach was successfully used to determine the properties of FCIs in quasi-1D geometry \cite{Johannes}, as well as in the thin-torus limit \cite{grusdt2014realization}. We use the Matrix Product State (MPS) formulation of DMRG implemented in the iTensor package \cite{itensor}. Periodic boundary conditions were implemented by introducing hopping between the first and the last site in the MPS. Fig. \ref{fig:dmrg}(a) shows the energies of the four lowest states for $U=1000$, $\tau_1=2, \tau_2=-0.4$ and varying system size with constant filling factor $\nu=1/3$. The calculation for each energy level was continued until the difference in energy between two sweeps was $10^{-9}$. We note that the convergence was slow, sometimes more than 100 sweeps were needed, even thought we accelerated the calculations by using the eigenstates for lower values of $U$ as an initial guess. We investigated also the convergence of energy as a function of bond dimension $\chi$ by increasing $\chi$ by 100 and setting the convergence criterion at $10^{-7}$. However, even after the convergence the energies fluctuated, so we estimate the accuracy of these results to be $10^{-6}$, which is still enough for the purpose of this work. To ensure that DMRG does not converge to higher excited states we calculated 16 or more excited states for  $\chi=200$ for each system size. Also, for small systems (up to $N_{\mathrm{part}}=7$), we compared the results with exact diagonalization ones and found agreement within the accuracy estimated above. The degeneracy splitting of the ground states decreases with growing system size. The energy gap to excited states $\Delta E \approx 0.44$ for the largest system with $N_{\mathrm{part}}=10$ particles, and seems to extrapolate to a finite value in an infinite system. 

Fig. \ref{fig:dmrg}(b) shows the low-energy spectrum of a $L=12$ system with $N_{\mathrm{part}}=4$ as a function of $\tau_2$, obtained using the exact-diagonalization method. We have chosen negative $\tau_2$, so that the model can be rotated into the two-orbital model with positive staggered potential $M'$. In our ED calculation, the Hamiltonian is diagonalized in subspaces with conserved momentum. The momenta of the three lowest states (red points) agree with the generalized Pauli principle for FCI. The energy splitting between these states grows with increasing magnitude of $\tau_2$ hopping, which leads to lowering of the gap between the third and fourth state. This is consistent with results in Ref. \citenum{guo2012fractional}, obtained for a finite interaction of different kind. It is also seen that the degeneracy of lowest excited state, existing for $\tau_2=0$, is split when $|\tau_2|$ is increased, which also leads to the decrease of the energy gap. The black line denotes the single particle topological phase transition between trivial ($|\tau_2|>|\tau_1|$) and nontrivial ($|\tau_2|<|\tau_1|$) phases (corresponding to trivial and nontrivial regions of the 2D two-orbital model). It can be seen that the energy gap between the three lowest states (red) and the rest of the spectrum (blue) remains open in the whole nontrivial region. It is small but finite even in the nontrivial region, which may be a result of finite-size effects. 

\begin{figure}
\includegraphics[width=\textwidth]{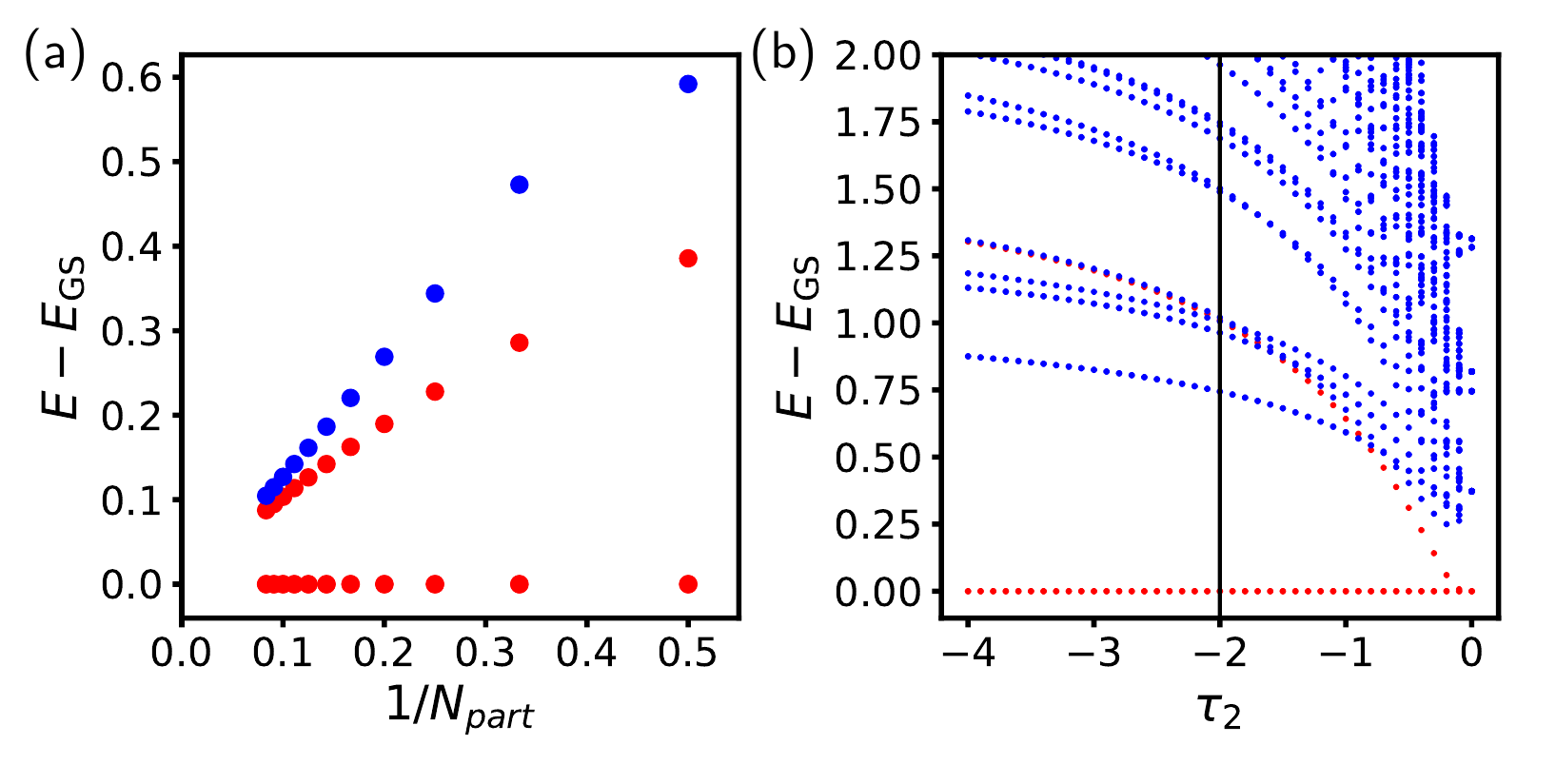}
\caption{Low-energy spectrum for the SSH model with $\nu=1/3$, $U=2$, $\tau_1=2$ and finite $\tau_2$: (a) energies of four lowest states for $\tau_2=-0.4$ as a function of inverse number of particles, obtained using DMRG method, (b) low-energy spectrum of $L=12$ chain with $N_{\mathrm{part}}=4$ particles, obtained using the exact diagonalization method. The red points in (a) mark the three lowest states. In (b), they mark the lowest states of momentum subspaces agreeing with the generalized Pauli principle for FCIs (being also the ground states of the whole system for small $\tau_2$). The single-particle topological phase transition is marked by a black vertical line in (b).}
\label{fig:dmrg2}
\end{figure}

\subsection{Finite $U$}
When $U$ is decreased to small finite values, Eq. \ref{eq:HX} is no longer valid, because the configuration $|4\rangle$ from Fig. \ref{fig:dimerized}(a) has to be included in the one-particle excitation. In the fully dimerized case, this leads to decrease of the energy gap, as seen in Fig. \ref{fig:dimerized}(b). Further lowering of the energy gap occurs due to the lifting of degeneracy of both ground states and excitations when finite $\tau_2$ is introduced.

Fig. \ref{fig:dmrg2}(a) shows the DMRG results for $U=2$, $\tau_1=2, \tau_2=-0.4$ and varying system size. Now the splitting between the three lowest states is much larger than for infinite $U$, although it may vanish in the thermodynamic limit. The energy gap is smaller than in the previous case. It does not close for all the system sizes we consider, but one cannot be certain whether it survives in the thermodynamic limit. We note that in $U=2$ case the convergence was much faster than for $U=1000$ and we were able to increase both system size and accuracy (now being $10^{-7}$ or more for small system sizes). At least four excited states were calculated for each system size. For $N_{\mathrm{part}}>10$ this number was larger to ensure that the states we capture are indeed the lowest.
 
Fig. \ref{fig:dmrg2}(b) shows the evolution of the energy spectrum for $L=12, N_{\mathrm{part}}=4$ with increasing $|\tau_2|$. Now, the energy gap closes within the nontrivial region. This may be a result of bandwidth ($2|\tau_2|$) becoming comparable with the interaction. 

\subsection{Effects of the staggered potential}\label{ssec:staggered}

When finite staggered potential $\varepsilon$ is introduced at $\tau_2=0$ the model is still dimerized, however the particle density within each dimer is no longer evenly distributed between the A, B orbitals (see Eq. \ref{eq:fullydimerized}). We parametrize the model with $\varepsilon=\tau\sin(\alpha), \tau_1=\tau\cos(\alpha)$. Increasing $\alpha$ from 0 to $\pi/4$ induces the transition between the fully dimerized model (nontrivial bands) and the model with isolated individual sites (trivial bands). In this parametrization, the single-particle band structure remains constant all through the transition, see Eq. \ref{eq:fullydimerized}. The nontrivial system can be transformed into a trivial gap without closing the band gap, because the staggered potential breaks the chiral symmetry which protects the topological phase.

The ground state is three-fold degenerate as in the $\varepsilon=0$ case, as long as $\alpha \neq \pi/4$. However, the excitation energy varies. In the infinite $U$ case, this can be seen from the effective Hamiltonian matrix describing one particle excitation, which, in analogy to Eq. \ref{eq:HX} has the form
$$
H_1=\tau\left[\begin{matrix}
-\sin(\alpha) & \cos(\alpha) & 0 \\
\cos(\alpha) & 0 & \cos(\alpha) \\
0 & \cos(\alpha)& \sin(\alpha) 
\end{matrix}\right],
$$
whose eigenvalues can be calculated exactly. The lowest one-particle excitation energy is then $$E^{\mathrm{X}}_1=\tau\left(2-\sqrt{2}\sqrt{\sin(\alpha)^2+1}\right),$$ 
which vanishes for $\alpha=\pi/4$. Hence, the gap closes when the system consists of isolated sites.

\begin{figure}
\includegraphics[width=\textwidth]{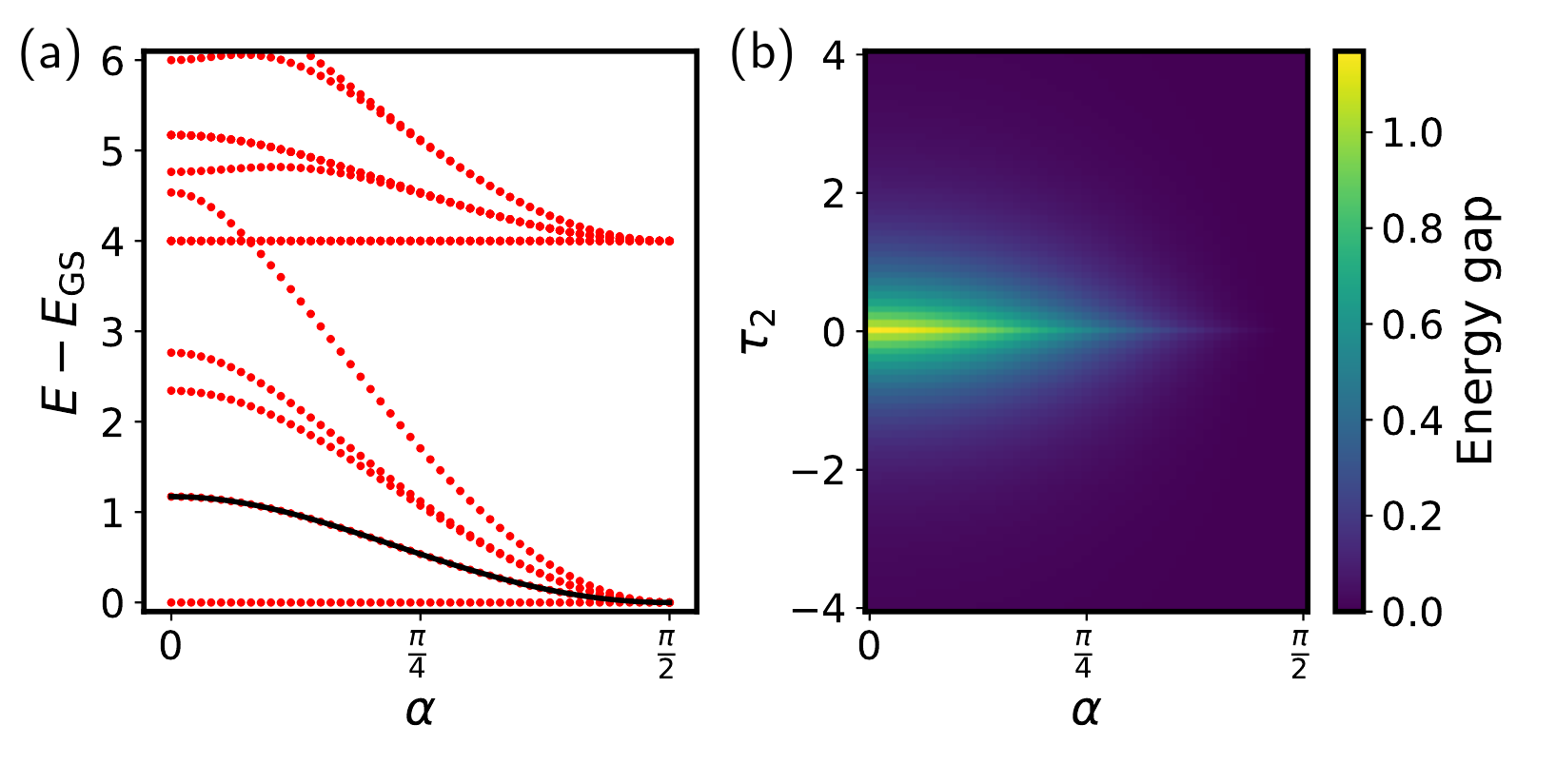}
\caption{Effect of the staggered onsite potential on $N_{\mathrm{part}}=4, L=12$ SSH chain with $\tau_1=2$ and $U=1000$: a) The energy spectrum in fully dimerized limit, as a function of parameter $\alpha$. b) the energy gap between the FCI-like ground states and rest of the spectrum, as a function of $\alpha$ and $t_2$. The black line in (a) denotes the analytical result for the one-particle excitation.}
\label{fig:staggered}
\end{figure}

In Fig. \ref{fig:staggered}(a) we show result for a system with $U=1000$, $\tau=2$ $N_{\mathrm{part}}=4, L=12$. $E_2$, indicated by a black solid line, is the lowest excitation energy in the whole range $\alpha \in [0,\pi/4)$, hence the gap remains open for any $\alpha\neq \pi/4$. If nonzero $\tau_2$ is added, its effect would be similar to those described in Subsection \ref{ssec:infinite} for $\varepsilon=0$ case: the removal of the degeneracy of ground states and of excitations. Fig. \ref{fig:staggered}(b) shows the the magnitude of the energy gap as a function of both $\alpha$ and $\tau_2$ for constant $U=1000$ and $\tau=2$ . It can be seen that the maximum is at $\tau_2=0$, $\alpha=0$, i.e. the fully dimerized system with no staggered potential.

These results show the importance of the dimer structure for stability of the FCI-like phase in our model. The deviations from the fully dimerized structure with $\varepsilon=0$, either in the form of inter-dimer coupling or staggered potential, lead to the decrease, and eventually the vanishing of the many-body gap. We note that this conclusion is valid for the interaction defined by Eq. \ref{eq:interaction}. For other kinds of interaction, the results may differ (see e.g. Ref. \citenum{budich2013fractional}). 

\subsection{The checkerboard lattice}

As a last point, let us show that the gap is stable in the infinite interaction limit also for the checkerboard model. We still treat the two orbitals of the unit cell as one site, and apply the interaction defined in Eq. \ref{eq:interaction}. This is different from the standard interaction used for checkerboard model. Usually one treats each orbital as separate site and applies the nearest-neighbor density-density interaction (see e.g. Ref. \citenum{Neupert}). Such an approach would be much more complicated because after rotation additional non-density-density terms will appear, and our construction of excited states would no longer be applicable.

\begin{figure}
\includegraphics[width=\textwidth]{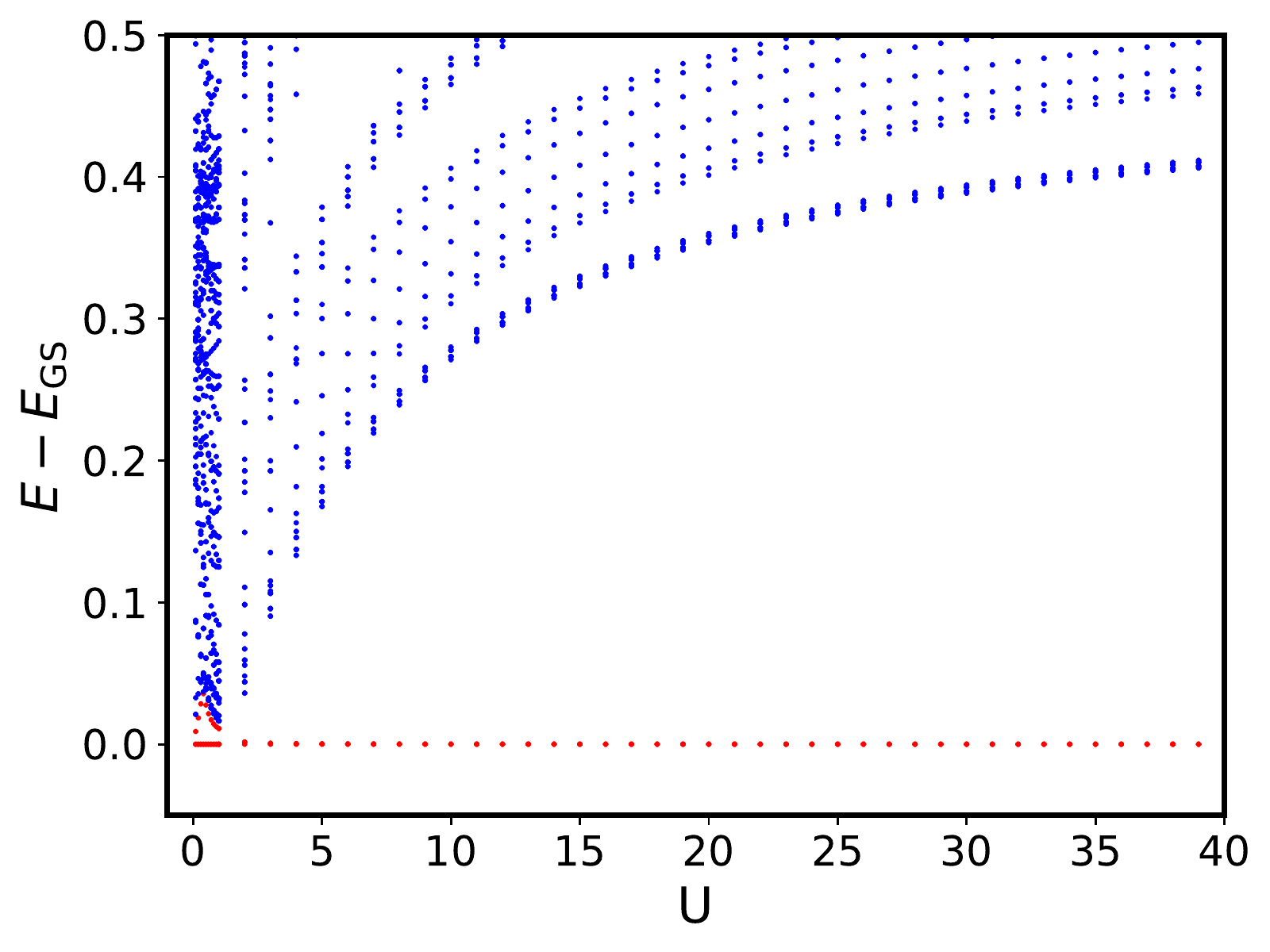}
\caption{The energy spectrum of the 1D checkerboard chain of length $L=12$ and $N_{\mathrm{part}}=4$,as a function of the interaction strength $U$. The single-particle parameters are $t_1=\sqrt{2} t_2=1$, $t_\varepsilon=t_3=0$ (see the \ref{sec:checkerboard} for their definitions). The red dots denote the three lowest states.}
\label{fig:checkerboard}
\end{figure}

Figure \ref{fig:checkerboard} shows the evolution of energy spectrum of the thin-torus limit of the checkerboard model as a function of $U=V$. Here, the parameters correspond to the nearly flat band case described in Ref. \citenum{Neupert}. The energy gap between three lowest states and the rest of spectrum tends to a finite value for $U\rightarrow \infty$, similarly to the one in Fig. \ref{fig:dimerized}. This result can be explained in terms of the dimer structure of the wavefunctions. It can be shown that the system whose spectrum is shown in Fig. \ref{fig:checkerboard} is a rotated SSH chain plus a small correction (see the \ref{sec:checkerboard} for details). The parameters of the SSH chain are $\alpha\approx 0.27 \pi$ and $\tau_2\approx 0.107\tau$, which corresponds to $\tau_2 \approx 0.213$ in Fig. \ref{fig:staggered}. This shows that, in comparison to the ideal $\alpha=0, \tau_2=0$ case, the gap is significantly lowered due to the staggered potential, while the effect of the inter-dimer hopping $\tau_2$ is much smaller. The correction introduces further inter-dimer hoppings (including a second-neighbor one), but they are still small compared to $\tau$, hence we can still attribute the stability of the FCI-like phase to the dimerization. A similar reasoning can be applied to the version of the checkerboard model presented in Ref. \citenum{Sun}, involving the third-neighbor hoppings  (see the \ref{sec:checkerboard} for details). Its validity is confirmed by the exact-diagonalization calculations, which yield the dependence of energy spectrum on $U$ similar to the one in Fig. \ref{fig:checkerboard}.

\section{Summary and conclusions}
Motivated by Ref. \citenum{FarExceed}, we studied the simplified version of the problem of stability of Fractional Chern Insulators with respect to interband excitations by considering the system in the thin-torus limit. We focused on the extended SSH model, related to the 1D limit of the two-orbital flat-band model by a basis rotation. For the fully dimerized case with no staggered potential, we have obtained analytical expressions for the excitation energies and shown that for infinite interaction the many-body energy gap between the ground state manifold and excited states remains finite even in the thermodynamic limit. Next, we investigated the effect of inter-dimer hopping and staggered potential. We have shown that both perturbations lower the many-particle gap, although it remains open if they are small enough. This indicates that the dimer structure is responsible for the stability of the FCI-like ground state. Finally, by considering the thin-torus limit of the checkerboard model, we show that the interpretation of the excitations in terms of dimerized wavefunctions is not restricted to one lattice model.

\section{Acknowledgement}
This work was supported by National Science Centre, Poland, grant PRELUDIUM no. 2016/21/N/ST3/00843.
\appendix

 \section{Relation to checkerboard model}\label{sec:checkerboard}
The 1D limit of the checkerboard limit is defined by the following hopping matrices
$$
\tilde{T}_{\mathrm{CB}}=\left[\begin{matrix}
t_2 & 0 \\
\sqrt{2} t_1 &  -t_2 
\end{matrix}\right],~\tilde{E}_{\mathrm{CB}}=\left[\begin{matrix}
-2t_2 & \sqrt{2} t_1 \\
\sqrt{2} t_1 &  2 t_2 
\end{matrix}\right].
$$
The notation is the same as in Ref. \citenum{Neupert} for 2D model. We set $t_1=\sqrt{2} t_2$ which guarantees the minimal flatness ratio (i.e. band dispersion divided by band gap)\cite{Neupert}. In such a case, the nearest-neighbor hopping matrix reads
$$
\tilde{T}_{\mathrm{CB}}=\tilde{T}_1+\tilde{T}_2, ~~\tilde{T}_1=\left[\begin{matrix}
t_2 & -t_\varepsilon \\
2 t_2 &  -t_2
\end{matrix}\right]~~\tilde{T}_2=\left[\begin{matrix}
0 & t_\varepsilon \\
0 &  0
\end{matrix}\right],$$
where $t_{\varepsilon}$ is a new term, corresponding to fourth-neighbor hopping in the 2D checkerboard model. We introduce it to divide the hopping matrix into the part which can be rotated into the SSH model ($\tilde{T}_1$) and the remaining corrections ($\tilde{T}_2$). If $t_\varepsilon=t_2/2$, $\tilde{T}_1$ is a hopping matrix of the SSH model with $\tau_1=5/2t_2$ rotated by an angle $\phi=\arcsin(-1/\sqrt{5})$. If we rotate the $\tilde{E}$ matrix back by this angle, it will become
$$
E_{\mathrm{CB}}=\left[\begin{matrix}
-\frac{14}{5}t_2 & -\frac{2}{5}t_2 \\
-\frac{2}{5}t_2 &  \frac{14}{5}t_2
\end{matrix}\right].
$$
Hence, if we neglect the presence of $\tilde{T}_2$, i.e. we consider the checkerboard model with additional hopping $t_\varepsilon$, we obtain a rotated SSH model. It is characterized by strong staggered potential $\varepsilon=\frac{14}{5}t_2$, relatively strong nearest-neighbor (intra-dimer) coupling $\tau_1=\frac{5}{2}t_2$ and relatively low inter dimer hopping $\tau_2=-\frac{2}{5}t_2$. Using the notation from Subsection \ref{ssec:staggered}, we have $\tau=\sqrt{\varepsilon^2+\tau_1^2}\approx 3.75 t_2$ and $\alpha=\arctan(\epsilon/\tau_2)\approx  0.27 \pi$.

For the exact checkerboard model, the neglected $\tilde{T}_2$ term has to be taken into account. After the rotation, it takes the form
$$
T_2=\left[\begin{matrix}
\frac{1}{5}t_2 & \frac{2}{5}t_2 \\
-\frac{1}{10}t_2 &  \frac{1}{5}t_2
\end{matrix}\right].
$$
It slightly reduces the intra-dimer hopping $\tau_1$, but also introduces new terms not present in SSH model, corresponding to additional couplings between nearest-neighbor dimers as well as additional second-neighbor inter-dimer couplings. Nevertheless, they are still small compared to $\tau$, hence the intra-dimer terms still dominate.

To further flatten the lower band, another term $t_3$ can be introduced, corresponding to the third-neighbor hopping in the original 2D model \cite{Sun}. The hopping matrices in such a case are given by
$$
\tilde{T}_{\mathrm{CB2}}=\left[\begin{matrix}
t_2+t_3 & 0 \\
\sqrt{2} t_1 &  -t_2+t_3 
\end{matrix}\right], $$
$$\tilde{E}_{\mathrm{CB2}}=\left[\begin{matrix}
-2t_2+2t_3 & \sqrt{2} t_1 \\
\sqrt{2} t_1 &  2 t_2+2t_3 
\end{matrix}\right].
$$
To minimize the flatness ratio, $t_2=\frac{t_1}{2+\sqrt{2}}$ and $t_3=\frac{t_1}{2+2\sqrt{2}}$ is used \cite{Sun}. Again, we can introduce $t_\varepsilon$ and divide the $\tilde{T}_{\mathrm{CB2}}$ into two parts. The first yields the SSH model rotated by $\arctan(1/(2\sqrt{2}+2))$, while the second contains small corrections.

The fact that $t_3$ flattens the lower band may be also understood in terms of dimer wavefunctions, by the means of the perturbation theory. Let us consider only the lowest band of a fully dimerized system and introduce the inter-dimer terms as a perturbation. If $t_2=\frac{t_1}{2+\sqrt{2}}, t_3=0$, the inter-dimer terms yield a negative inter-dimer coupling. Since the $t_3$ terms are proportional to unit matrix, they are not affected by the rotation. As a result, the contribution of $t_3$ is positive, so the effective inter-dimer hopping amplitude is lowered and the band gets flattened. On the other hand, in the upper band both contributions are positive, hence the addition of $t_3$ increases the band dispersion.

\bibliographystyle{elsarticle-num}
\bibliography{thintorus}
\end{document}